\documentclass[
    reprint, 
    superscriptaddress,
    amsmath,amssymb,
    aps,
    prx,
    floatfix,
]{revtex4-2}

\usepackage{graphicx}
\usepackage{xcolor}
\usepackage{siunitx}
\usepackage{braket}
\usepackage{bm}
\usepackage[colorlinks=true,linkcolor=blue,citecolor=blue]{hyperref}
\usepackage[capitalise]{cleveref}

\begin{document}
\title{A high-efficiency plug-and-play superconducting qubit network}

\author{Michael Mollenhauer}
\thanks{These authors contributed equally.}
\author{Abdullah Irfan}
\thanks{These authors contributed equally.}
\author{Xi Cao}
\thanks{These authors contributed equally.}
\author{Supriya Mandal}
\affiliation{Department of Physics, University of Illinois at Urbana-Champaign, Urbana, IL 61801, USA}
\author{Wolfgang Pfaff}
\email{wpfaff@illinois.edu}
\affiliation{Department of Physics, University of Illinois at Urbana-Champaign, Urbana, IL 61801, USA}
\affiliation{Materials Research Laboratory, University of Illinois at Urbana-Champaign, Urbana, IL 61801, USA}

\begin{abstract}
    Modular architectures are a promising approach to scale quantum devices to the point of fault tolerance and utility \cite{kimble_quantum_2008,monroe_largescale_2014,awschalomDevelopmentQuantumInterconnects2021}.
    Modularity is particularly appealing for superconducting qubits, as monolithically manufactured devices are limited in both system size and quality \cite{angArchitecturesMultinodeSuperconducting2022,bravyiFutureQuantumComputing2022,smithScalingSuperconductingQuantum2022}.
    Constructing complex quantum systems as networks of interchangeable modules can overcome this challenge through `Lego-like' assembly, reconfiguration, and expansion, in a spirit similar to modern classical computers.
    First prototypical superconducting quantum device networks have been demonstrated \cite{rochObservationMeasurementInducedEntanglement2014,dickelChiptochipEntanglementTransmon2018,narlaRobustConcurrentRemote2016,leungDeterministicBidirectionalCommunication2019b, zhongDeterministicMultiqubitEntanglement2021, niuLowlossInterconnectsModular2023a, qiuDeterministicQuantumTeleportation2023,burkhartErrorDetectedStateTransfer2021a,kurpiersDeterministicQuantumState2018b, campagne-ibarcqDeterministicRemoteEntanglement2018b, axlineOndemandQuantumState2018b,zhouRealizingAlltoallCouplings2023a}. 
    Interfaces that simultaneously permit interchangeability and high-fidelity operations remain a crucial challenge, however.
    Here, we demonstrate a high-efficiency interconnect based on a detachable cable between superconducting qubit devices.
    We overcome the inevitable loss in a detachable connection through a fast pump scheme, enabling inter-module SWAP efficiencies at the 99\%-level in less than $\qty{100}{\nano\second}$.
    We use this scheme to generate high-fidelity entanglement and operate a distributed logical dual-rail qubit.
    At the observed $\sim$1\% error rate, operations through the interconnect are at the threshold for fault-tolerance.
    These results introduce a modular architecture for scaling quantum processors with reconfigurable and expandable networks.
\end{abstract}

\maketitle

\section{Introduction}

\begin{figure*}[bt]
\centering
\includegraphics{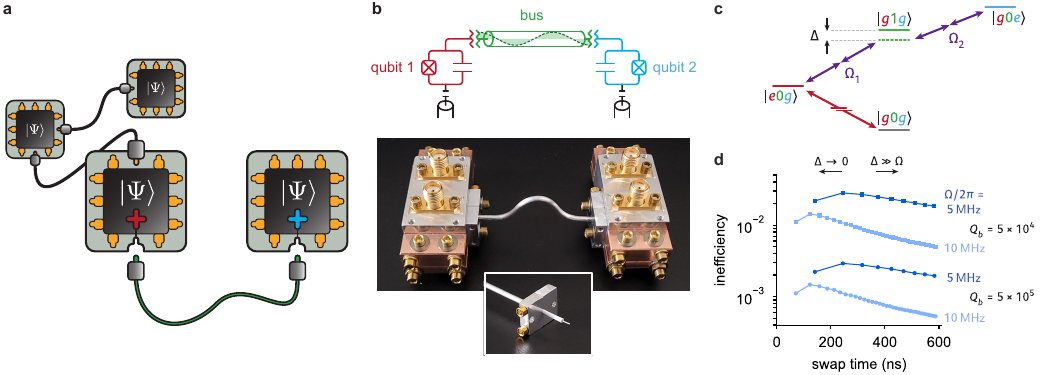}
\caption{\textbf{Experiment overview.}
    \textbf{a,} Scaling quantum devices as a network of modules permits reconfiguration or replacement of subsystems.
    The minimal requirement is the ability to transmit quantum states with high fidelity between qubits in different modules (shown as blue and red crosses) through detachable links (shown as green cable).
    \textbf{b,} Experimental implementation of a `plug-and-play' quantum network.
    A superconducting coaxial cable with custom connectors couples capacitively to transmon qubits and acts as a modular, high-$Q$ quantum bus.
    Here: a $\qty{73}{\milli\meter}$ cable, with the $2\lambda$ mode acting as the bus.
    Both qubits and their readout resonators are on separate chips, packaged in separate housings.
    \textbf{c,} Sketch of the level diagram and drive scheme for Raman transitions through the bus.
    Pumping each transmon separately realizes two-photon sideband transitions (purple arrows) between qubits (red and blue levels) and bus mode (green).
    Qubit-qubit swaps are realized for $\Omega_1 = \Omega_2 \equiv \Omega$.
    \textbf{d,} Simulated inefficiency of a swap from qubit~1 to qubit~2 as a function of the swap time, assuming ideal qubits.
    Squares and circles: $Q_b = 50,000$ (as shown for NbTi cables \cite{burkhartErrorDetectedStateTransfer2021a}) and $Q_b = 500,000$ (Al cables, see \cref{app: setup}), respectively.
    Dark and light curves: $\Omega/2\pi = \qty{5}{\mega\hertz}$ and $\qty{10}{\mega\hertz}$, respectively.
    Shorter times are realized with small detunings (leftmost points: $\Delta = 0$), longer times are in the detuned Raman regime.
}
\label{fig:fig1}
\end{figure*}

At the core of a modular device architecture lies the interchangeability of subsystems. 
Assembly from interchangeable components is routine in classical computers, and has been recognized also as a key scaling approach for quantum processors \cite{kimble_quantum_2008, monroe_largescale_2014,awschalomDevelopmentQuantumInterconnects2021, angArchitecturesMultinodeSuperconducting2022}.
For one, seamless addition and removal of components enables system upgrades with pre-tested, higher-fidelity qubit modules.
Second, expansion of system size and computational power can be achieved simply by plugging in additional modules (\cref{fig:fig1}a).
Because manufactured devices inevitably have finite yield and fluctuating parameters, it is difficult to realize monolithic quantum systems containing large numbers of high-quality qubits;
This is a particularly important challenge in solid-state systems, such as superconducting qubits. 
Modular quantum computing architectures can thus be key for overcoming a central challenge in realizing large-scale quantum processors
\cite{angArchitecturesMultinodeSuperconducting2022,bravyiFutureQuantumComputing2022,smithScalingSuperconductingQuantum2022}.

Achieving this vision of modular scaling hinges critically on efficient interchangeable interconnects.
Previous works have accepted compromises in network capability or performance due to the difficulty in combining low loss and interchangeability.
Recent examples include tolerating sparse weak links \cite{rametteFaulttolerantConnectionErrorcorrected2024}; performing measurement-based protocols in which entanglement is realized non-deterministically \cite{rochObservationMeasurementInducedEntanglement2014, dickelChiptochipEntanglementTransmon2018, narlaRobustConcurrentRemote2016}; or achieving low loss by sacrificing plug-and-play interchangeability using wafer- or wire-bonding \cite{goldEntanglementSeparateSilicon2021,conner_superconducting_2021,zhongDeterministicMultiqubitEntanglement2021,niuLowlossInterconnectsModular2023a,qiuDeterministicQuantumTeleportation2023}.
In contrast, fully embracing modularity as a key scaling strategy requires inter-module links on par with intra-module gates, with deterministic operations at the 1\% error level \cite{fowlerHighthresholdUniversalQuantum2009,barendsSuperconductingQuantumCircuits2014a} between interchangeable modules.
Losses in experiments to date exceed 15\% \cite{axlineOndemandQuantumState2018b,campagne-ibarcqDeterministicRemoteEntanglement2018b,kurpiersDeterministicQuantumState2018b,burkhartErrorDetectedStateTransfer2021a,zhouRealizingAlltoallCouplings2023a}, however, well in excess of requirements for scaling.

Here, we introduce a high-efficiency modular architecture for interchangeable superconducting quantum devices.
We have developed an interconnect between separately packaged transmon qubits that combines a reliably (un)pluggable coaxial-cable link with fast and high-fidelity Raman transitions.
Using this architecture, we demonstrate inter-module SWAP gates with $\sim 1\%$ loss in under $\qty{100}{\nano\second}$.
Remarkably, our interconnect rivals the performance shown with superconducting bonds between qubit circuits and cables \cite{zhongDeterministicMultiqubitEntanglement2021,niuLowlossInterconnectsModular2023a,qiuDeterministicQuantumTeleportation2023}.
The speed and efficiency of our pump scheme further enables the generation of high-fidelity inter-module entanglement, and the operation of a distributed dual-rail qubit.
Our interconnect does not require any circuit elements beyond the intrinsic nonlinearity of the qubits, making it applicable for different types of qubits beyond the transmon.

\section{Interconnect design}

A quantum interconnect scheme should combine low loss with flexibility in assembly and connectivity.
The latter makes cable-based connections highly attractive.
Additionally, superconducting coaxial cables can be used as a low-loss transmission line resonator \cite{burkhartErrorDetectedStateTransfer2021a,zhongDeterministicMultiqubitEntanglement2021,leungDeterministicBidirectionalCommunication2019b,niuLowlossInterconnectsModular2023a,qiuDeterministicQuantumTeleportation2023,kurpiers_characterizing_2017}, allowing their use as an off-chip version of a quantum bus \cite{majerCouplingSuperconductingQubits2007b}.
Realizing a detachable connection between qubit and cable implies a capacitive coupling scheme as illustrated in \Cref{fig:fig1}b.
To achieve reliable yet interchangeable connectivity, we have designed a connector mechanism that yields quality factors for a demountable bus made of Al cable of up to $Q_b \approx 500,000$ (\cref{fig:wiring diagram and device}c).
Here, we have mounted the connector to a sub-cutoff waveguide containing a fixed-frequency transmon \cite{axlineArchitectureIntegratingPlanar2016}. 
This design is readily compatible with leading, hardware-efficient approaches to quantum error correction based on high-$Q$ cavities \cite{ gertler_protecting_2021, chakram_multimode_2022, sivakRealtimeQuantumError2023a, ni_beating_2023, milulSuperconductingCavityQubit2023, koottandavidaErasureDetectionDualRail2024,ganjamSurpassingMillisecondCoherence2024}, for which a modular scaling paradigm is particularly attractive \cite{chouDeterministicTeleportationQuantum2018}.

To complete the interconnect, the physical link must be combined with a compatible, high-fidelity gate scheme.
The capacitive coupling between cable and qubit comes with two important consequences we need to consider:
First, the attainable cable $Q$ is reduced compared to what has been shown with galvanically bonded cable connections; second, it precludes the use of fast tunable couplers based on flux-tuning \cite{zhongDeterministicMultiqubitEntanglement2021, niuLowlossInterconnectsModular2023a,qiuDeterministicQuantumTeleportation2023}.
To realize a sufficiently fast gate, we target a Raman-type process \cite{burkhartErrorDetectedStateTransfer2021a} by parametrically driving sideband transitions between the qubits and bus mode (\cref{fig:fig1}c).
Fixed-frequency transmon qubits allow for driving such transitions natively \cite{wallraffSidebandTransitionsTwoTone2007}.
In order to realize gate speeds that strongly exceed interconnect loss rates we have designed a low-frequency driving scheme that enables fast sideband rates without introducing excess decoherence (see below, and \cref{app: sideband model}).

To assess the performance of the interconnect scheme we consider the efficiency of a SWAP gate between distant qubits. 
Executing high-fidelity SWAPs is sufficient for executing inter-module circuits in a network of multi-qubit modules \cite{zhongDeterministicMultiqubitEntanglement2021}.
The performance of the targeted SWAP is determined by the following key factors.
Loss in the bus leads to inefficiency that can be counteracted by introducing a detuning \cite{bergmannCoherentPopulationTransfer1998}. 
This comes at the cost of slower speeds, which in turn leads to inefficiency due to finite qubit coherence.
\Cref{fig:fig1}d shows predictions for how $\sim 1\%$-level SWAP inefficiency can be reached (\cref{app: swapp efficiency}), leading us to the following strategy.
Generally, $\Omega$ should be as high as possible. 
Because we must anticipate that cable coupling has a detrimental effect on the qubit, we prioritize speed to avoid inefficiency from qubit decoherence.
As shown in \Cref{fig:fig1}d, high $Q_b$ and fast sidebands predict excitation swaps with sub-percent loss in around $\qty{100}{\nano\second}$, much faster than currently achievable transmon coherence times.

\section{Qubit-bus sidebands}

\begin{figure}
\includegraphics{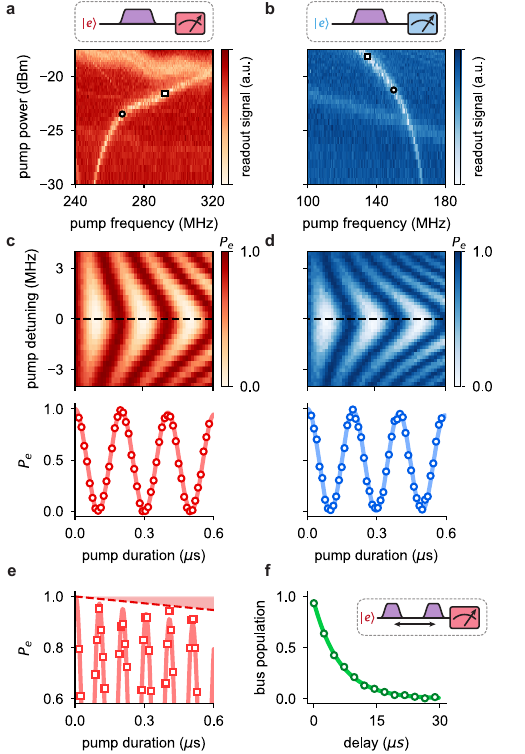}
	\caption{
    \textbf{Fast sidebands between qubit and high-$Q$ detachable bus.}
    \textbf{a, b,} Readout signal of the left (red) and right (blue) qubits as a function of  pump frequency and device input power.  
    Circle and square markers along the sideband resonance indicate $\Omega_i/2\pi = \qty{5}{\mega\hertz}$ and $\qty{10}{\mega\hertz}$, respectively.
    \textbf{c, d,} Top: Time evolution of the excited state population, $P_\text{e}$, of qubits~1 (red) and 2 (blue) during pumping with $\Omega_i/2\pi = \qty{5}{\mega\hertz}$.
    Bottom: Linecuts at zero detuning.
    Lines are a model taking into account $\Omega_i$ and undriven decoherence rates of qubit and bus mode.
    \textbf{e,} Excited state population from a fit to sideband oscillations measured on qubit 1 with $\Omega_1/2\pi = \qty{10}{\mega\hertz}$.
    Red shaded region shows the limit based on undriven decoherence rates.
    \textbf{f,} Measurement of the single-photon decay of the bus mode.
    Fit to an exponential decay yields a lifetime of $\tau = \qty{6.2}{\micro\second}$.
    }
\label{fig:fig2}
\end{figure}

At the heart of the targeted Raman process are sideband transitions that convert qubit excitations into bus photons, and that can be tuned up individually for each module.
To realize these transitions, the connector is designed such that a mode of the bus resonator is coupled to both qubits in the dispersive regime \cite{schusterResolvingPhotonNumber2007}. 
Sidebands can then be driven by applying a pump tone to the transmon, with a frequency
$\omega_i = (|\omega_b - \omega_{a,i}|)/2$, where $\omega_b$ and $\omega_{a,i}$ are the resonant frequencies of target bus mode and qubit $i$, respectively.
This can be understood as degenerate four-wave mixing, for which we predict sidebands that are faster than the more commonly employed non-degenerate case (\cref{app: sideband model}).
The system is designed such that $\omega_i$ is placed several GHz below transmon and bus (\cref{app: setup}) and thus strongly detuned from any excitable mode.
Near resonance, the Hamiltonian describing the process is $H_{\text{sb},i} = (\Omega_i/2) (e^{\mathrm{i} \Delta t}\hat{a}_i^\dag \hat{b} + \text{h.c.})$, where $\hat{a}$ and $\hat{b}$ are the annihilation operators for transmon $i$ and bus mode, and the sideband rate $\Omega_i$ is set by the pump strength.

We begin by calibrating the sidebands for both qubits individually.
To find $\omega_i$, we first prepared each qubit in the excited state and recorded its change in state as a function of the frequency and power after applying a pump tone of fixed length (\cref{fig:fig2}a,b).
Tuning the pump to resonance and compensating for Stark shifts (\cref{app: sideband tuneup}), rapid oscillations between the $\ket{e0}$ and $\ket{g1}$ states are induced, where the oscillation frequency $\Omega$ is set by the pump power.
\Cref{fig:fig2}c,d shows `sideband Rabi' oscillations for the case where $\Omega_{1,2}/2\pi$ is tuned to $\qty{5}{\mega\hertz}$. 
Importantly, we find that the decay of these oscillations can be explained by the undriven loss rates of qubit and cable alone, indicating that the strong pump does not cause excess decoherence.

Our simple model predicts that the sideband rate $\Omega$ can be increased simply by increasing pump power, thus speeding up the transition.
Larger rates can, however, come at the cost of additional loss.
As an example, \Cref{fig:fig2}e shows $\qty{10}{\mega\hertz}$ oscillations obtained for qubit~1.
Their decay is visibly faster than predictions from undriven loss rates, obtained from qubit coherence as well as the in-situ determined bus loss (\cref{fig:fig2}f). 
The onset of excess loss varies between devices, suggesting that mechanisms such as drive-induced leakage \cite{shillitoDynamicsTransmonIonization2022} or coupling to random two-level fluctuators \cite{abdurakhimovIdentificationDifferentTypes2022} play a role in setting the speed limit. 
In our experiment we have been able to reliably obtain `clean' $\qty{5}{\mega\hertz}$ oscillations; in the following we show that they enable low-loss inter-module excitation transfer.

\section{Inter-device gates}

\begin{figure}
    \includegraphics{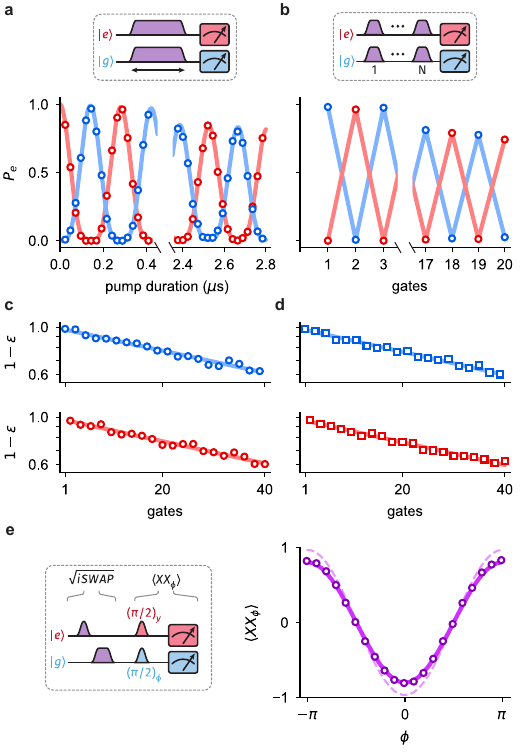}
        \caption{
        \textbf{Inter-module swaps.}
        \textbf{a,} Measured excited state populations after applying the pump for a variable time; $\Omega/2\pi = \qty{5}{\mega\hertz}$.  
        Solid line: model based on independently determined decoherence properties of (undriven) qubits and bus.
        \textbf{b,} Measured excited state population afer applying discrete SWAP gates.
        \textbf{c, d,} Total accumulated error as a function of number of SWAPs on qubits~1 (red) and 2 (blue) for $\Omega/2\pi = \qty{5}{\mega\hertz}$ and gate time $\qty{142}{\nano\second}$ (left panel) and $\Omega/2\pi = \qty{10}{\mega\hertz}$ and gate time $\qty{70}{\nano\second}$ (right panel).  
        \textbf{e,} Inter-module entanglement realized in a stroboscopic approach, verified by measuring $\langle XX_\phi \rangle$.
        Dots: raw data, not corrected for state preparation and measurement (SPAM) errors.
        Dashed line: simulation based on independently determined decoherence rates, not taking SPAM errors into account.
        Solid line: simulation with SPAM errors.
        Without SPAM, computed Bell state fidelity is 0.974.
        }
    \label{fig:fig3}
    \end{figure}

Having tuned up individual sidebands, we have next turned them on simultaneously to realize a Raman transition between the qubits.
Our sideband rates and bus-$Q$ predict a fast and high-efficiency SWAP for $\Delta=0$ (\cref{fig:fig1}d). 
From the in-situ determined $Q_b = 2.0\times10^5$ (\cref{fig:fig2}f) we predict a 0.97\% swap inefficiency for $\Omega_{1,2}/2\pi = \qty{5}{\mega\hertz}$, taking into account bus loss as well as qubit decoherence.
To test this prediction, we have prepared qubit 1 in $\ket{e}$, and applied both pumps at the same time.
After compensating for small inter-chip Stark-shifts (\cref{app: resonant Raman}), we have observed high-visibility oscillations in agreement with dynamics predicted by the total Hamiltonian, $H = H_{\mathrm{sb},1} + H_{\mathrm{sb},2}$, and established decoherence rates (\cref{fig:fig3}a).
The oscillation period is, as expected, $2\sqrt{2}\pi/\Omega$, and fitting the dynamics yields a loss per swap of $1.0(1)\%$, in quantitative agreement with our model (\cref{app: resonant Raman}).

To benchmark the achievable loss per gate, we have performed pulsed SWAP gates.
Because we drive sidebands with a second-order process, this mode of operation requires care in the pulse shaping to avoid excess errors (\cref{app: resonant Raman}) \cite{xiaFastSuperconductingQubit2023}.
Applying smoothed flat-top pulses with a length of 142\,ns, we have been able to achieve stroboscopic excitation swaps that match the quality of the continuous evolution (\cref{fig:fig3}b).
The loss per gate can be inferred by fitting the accumulated error as a function of number of swaps to an exponential decay (\cref{fig:fig3}c), yielding a loss rate of $1.2(1)\%$ per SWAP-gate; the excess error can be explained by tune-up imperfections (\cref{fig:qubit-qubit dynamics} and \cref{fig:qubit-qubit gate tune up}).
We have repeated the same procedure also with the sidebands tuned to $\Omega_{1,2}/2\pi = \qty{10}{\mega\hertz}$; while the gate time drops to 70\,ns in this case, the loss rate remains the same due to drive-induced excess errors discussed above (\cref{fig:fig3}d).
This inefficiency of order 1\% in a sub-100\,ns SWAP gate is a central benchmark result of our experiment, demonstrating a high-efficiency inter-device link.

Besides transferring population, the Raman transition needs to preserve phase coherence.
From the near-perfect agreement of time-dynamics with our model (\cref{fig:qubit-qubit dynamics}), coherence is already evident.
It can be shown more explicitly, however, by measuring transversal correlations after generating entanglement using the Raman process.
Fast sidebands and the high $Q$ of the bus enable us to generate inter-module entanglement in a simple, stroboscopic fashion.
We have first prepared qubit 1 in $\ket{e}$, then swapped half the population into the bus, followed by swapping the full bus population to qubit 2.
This routine corresponds to performing a $\sqrt{\mathrm{iSWAP}}$ entangling gate.
To analyze the expected entangled state, $(\ket{eg}+\ket{ge})/\sqrt{2}$, we have measured the correlator $\langle XX_\phi \rangle$, where $X_\phi$ is an operator rotated by $\phi$ with respect to $X$ (\cref{fig:fig3}e).
The data is in quantitative agreement with a simulation that only takes into account the undriven loss of all modes, predicting a Bell state fidelity of 97.4\%.
In this approach, the bus is populated for a significant time, explaining excess loss compared to the SWAP; this loss could, however, be suppressed with detuned Raman transitions.

\section{Detuned Raman regime}

\begin{figure*}
\includegraphics{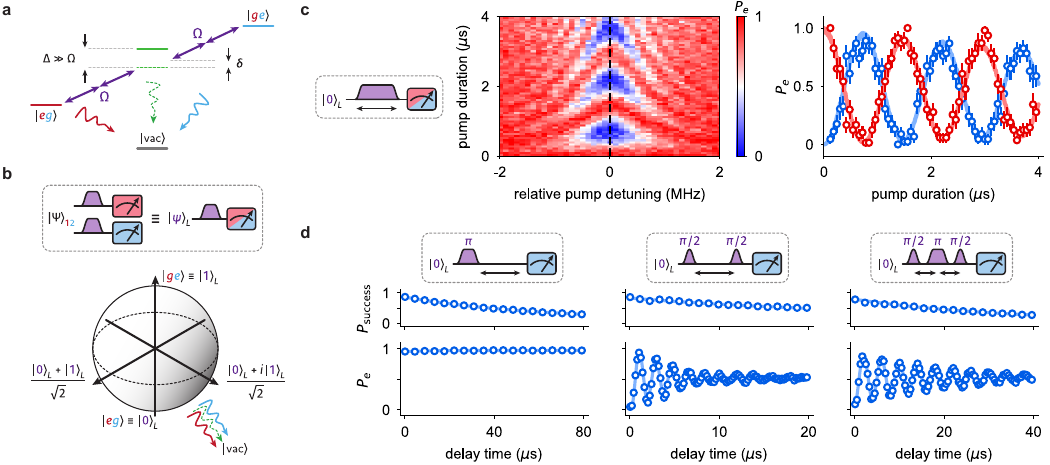}
	\caption{
    \textbf{Dual-rail operations in the detuned Raman regime.}
    \textbf{a,} Energy level schematic.
    Sidebands are applied off-resonantly w.r.t.~the bus mode and may be relatively detuned by $\delta$.
    Loss from the bus (green dashed arrow) can be suppressed at the cost of speed.
    \textbf{b,} Bloch sphere representation of the dual rail encoding.
    Photon loss relaxes the system into the vacuum, outside the qubit subspace.
    \textbf{c,} Time evolution of the qubit excited state populations (red: qubit~1; blue: qubit~2).
    Right panel: line cut at zero relative detuning.
    Line: model taking into account independently characterized decoherence parameters; the only free parameter is $\Omega$.
    \textbf{d,} Basic operations in the dual-rail subspace.
    After each sequence, both qubits are measured; events where both outcomes are $g$ are discarded.
    Top panels: fraction of measurement results, $P_\text{success}$, that were not discarded. 
    Bottom panels:
    Logical $T_1, T_{2\text{R}}$ and $T_{2\text{E}}$ measurements (left to right), post-selected on success.
    Data points shown are probabilities of qubit~2 being excited; lines are fits.
    $T_1$ is, as expected for a dual-rail encoding \cite{chouDemonstratingSuperconductingDualrail2023}, practically infinite.
    $T_2$ is limited by the individual qubits' coherence times.
    }
\label{fig:fig4}
\end{figure*}

During a Raman transition with $\Delta \gg \Omega_{1,2}$ the bus is populated only `virtually' and bus loss is suppressed (\cref{fig:fig4}a) \cite{bergmannCoherentPopulationTransfer1998}.
The coherent evolution of an excitation occurs then in the two-dimensional subspace spanned by the states $\ket{eg}$ and $\ket{ge}$; the rf-driven sidebands realize arbitrary rotations of this effective two-level system.
This amounts to operating a distributed dual-rail qubit (\cref{fig:fig4}b) \cite{teohDualrailEncodingSuperconducting2023} --- a logical  encoding that has recently garnered attention in the context of hardware-efficient quantum error correction with superconducting circuits \cite{levineDemonstratingLongCoherenceDualRail2024, koottandavidaErasureDetectionDualRail2024, chouDemonstratingSuperconductingDualrail2023}.
Reaching this detuned regime requires sidebands that are fast compared to loss rates.
The combination of low-loss interconnect and fast sideband transitions uniquely allows us to access this regime in a modularly connected network, providing a path toward loss-resilient entanglement and the operation of distributed logical qubits.

To demonstrate the capability to control the network in this fashion, we have first realized Rabi oscillations in the subspace $\{\ket{0}_L \equiv \ket{eg}, \ket{1}_L \equiv \ket{ge}\}$.
Following preparation in $\ket{eg}$, we have applied detuned sideband drives with $\Delta/\Omega \approx 5$.
In this regime, the time evolution of the system is governed by the effective Hamiltonian $H \approx (\Omega_R/2) (\hat{a}^{\dagger}_1 \hat{a}_2 e^{\mathrm{i}\delta t} + \text{h.c.})$, predicting oscillations with Rabi frequency $\Omega_R = \Omega^2 (2\Delta - \delta)/2 [ \Delta(\Delta - \delta)]$, where $\delta$ is a relative detuning between the pumps (\cref{fig:fig4}a).
\Cref{fig:fig4}c shows the characteristic chevron pattern obtained by measuring the time-dependent populations of both qubits, showing dynamics that are in quantitative agreement with numerical predictions.
Importantly, we have observed oscillations that are still much faster than effective loss rates.
The swap period shown here is, despite operating in a strongly detuned regime, on the sub-microsecond scale.
At the half-swap time, corresponding to a $\pi/2$ pulse of the dual-rail qubit, the two physical qubits are in a maximally entangled Bell state.
From the simulated time dynamics we estimate a Bell state fidelity of 0.98(1).
While the effect of loss in the bus is suppressed compared to the resonant regime, fidelity is now limited by on-site decoherence.

Having observed high-contrast oscillations in the dual-rail subspace, we can next perform logical operations.
Universal rotations can be executed by applying both pumps simultaneously with $\delta = 0$ while controlling duration and phase.
We demonstrate this in a proof-of-concept fashion by measuring the elementary coherence properties of the dual-rail qubit (\cref{fig:fig4}d).
A central feature of dual-rail qubits are erasure errors, as population loss relaxes the system to the vacuum outside the logical subspace.
Here, we simply detect these erasures at the end of each experiment and eliminate them by postselection \cite{chouDemonstratingSuperconductingDualrail2023}.
Our hardware could, however, be extended to perform mid-circuit erasure checks \cite{levineDemonstratingLongCoherenceDualRail2024,degraafMidcircuitErasureCheck2024} that would be required in an error correction code.
While our simple demonstration does not have the biased errors required to lower the overhead for error correction \cite{kubicaErasureQubitsOvercoming2023}, it provides an interesting perspective for logical qubit encodings in a distributed setting, and illustrates the potential for suppressing interconnect loss.

\section{Perspectives}

In summary, we have demonstrated a prototypical quantum network of interchangeable superconducting quantum devices.
We have developed a high-$Q$ detachable quantum bus connection and combined it with a low frequency parametric pumping scheme to realize a high-efficiency interconnect at the 1\,\% error level with operations on the order of 100\,ns.  
Our drive scheme further enables the generation of high-fidelity entanglement and the operation of a distributed logical qubit.
For practical reasons, we have employed a fairly short cable in our experiment. 
We note, however, that as long as $\Omega$ is small compared to the free spectral range of the cable, no detrimental effects are expected, placing the length limit at the one-meter scale.
Crucially, we find that the interconnect can be (re)attached with reproducible performance.
System parameters fluctuate strongly between assembly cycles, which can be expected due to the simplicity of the mounting system;
we have found, however, that fast sidebands can be achieved reliably (\cref{fig:repeatability test}).

While the demonstrated performance and robustness of the interconnect is already highly promising for modular scaling, we anticipate that substantial improvements are possible by improving the mechanical design.
The presented version of the connector hardware has a noticeable impact on qubit coherence.
A more mature design that preserves state-of-the-art coherence times \cite{placeNewMaterialPlatform2021,somoroff_millisecond_2023,wang_practical_2022,kono_mechanically_2024,ganjamSurpassingMillisecondCoherence2024} would enable intermodule gate infidelities approaching $10^{-3}$ (\cref{fig:fig1}d).
In this limit, operating on inter-module logical qubits would be on equal footing as intra-module operations, obviating the need for considering interconnects as weak links.

Our results pave the way for modular scaling of quantum processors. 
Transferring the interconnect hardware to more complex modules is readily achievable with existing technology.
Importantly, the plug-and-play nature of our interconnect enables incremental changes to the network. 
This ability allows processor nodes to be added or replaced over time without compromising existing parts.
The hardware design used in our work readily enables modular scaling of leading architectures for fault-tolerant quantum computing with bosonic encodings in high-$Q$ cavities \cite{ gertler_protecting_2021, chakram_multimode_2022, sivakRealtimeQuantumError2023a, ni_beating_2023, milulSuperconductingCavityQubit2023, koottandavidaErasureDetectionDualRail2024,ganjamSurpassingMillisecondCoherence2024}.
While we have chosen a stripline-based architecture in our experiment, a qubit-to-cable coupling compatible with coplanar-waveguide (CPW) based processors may be realized by introducing antenna transitions \cite{safwatNovelTransitionDifferent2002, beereshaCPWMicrostripTransition2016}.

Finally, the simplicity of our interconnect scheme, not requiring specific coupler circuitry, makes it also appealing beyond transmon qubits.
The only requirement for driving the sideband transitions is a suitable nonlinearity which can also be realized with other types of superconducting qubits such as the fluxonium \cite{vool_driving_2018,nie_parametricallycontrolled_2024}.
Additionally, we can envision high-fidelity interconnects between superconducting qubits and modalities that couple to superconducting resonators but have different environmental or packaging demands, such as spins \cite{lachance-quirionEntanglementbasedSingleshotDetection2020,harvey-collard_coherent_2022}.
Our results thus establish a modular architecture for scalable homogeneous and heterogeneous quantum device networks.

\section*{Acknowledgements}
The research was carried out in part in the Materials Research Lab Central Facilities and the Holonyak Micro and Nanotechnology Lab at the  University of Illinois.
We thank K.~Chow and R.~Goncalves for help with fabrication; and B.~DeMarco and A.~Kou for critical reading of the manuscript.
We acknowledge funding from the NSF Quantum Leap Challenge Institute for Hybrid Quantum Architectures and Networks (Award 2016136), from the IBM-Illinois Discovery Accelerator Institute, and the Army Research Office (Grant Number W911NF-23-1-0096). The views and conclusions contained in this document are those of the authors and should not be interpreted as representing the official policies, either expressed or implied, of the Army Research Office or the U.S. Government.

\section*{Author Contributions}
M.M. designed the device.  
M.M., A.I., and X.C. developed the theory models, conducted the experiment, and analyzed the data.  
S.M., M.M. and A.I. developed and tested the sideband pumping scheme.  
The paper was written by M.M., A.I., X.C., and W.P., with comments from all authors.  
The work was conceived and supervised by W.P.

\appendix

\section{Experimental methods}
\label{app: experimental methods}

\subsection{Setup}
\label{app: setup}

\begin{figure*}[pt]
\includegraphics{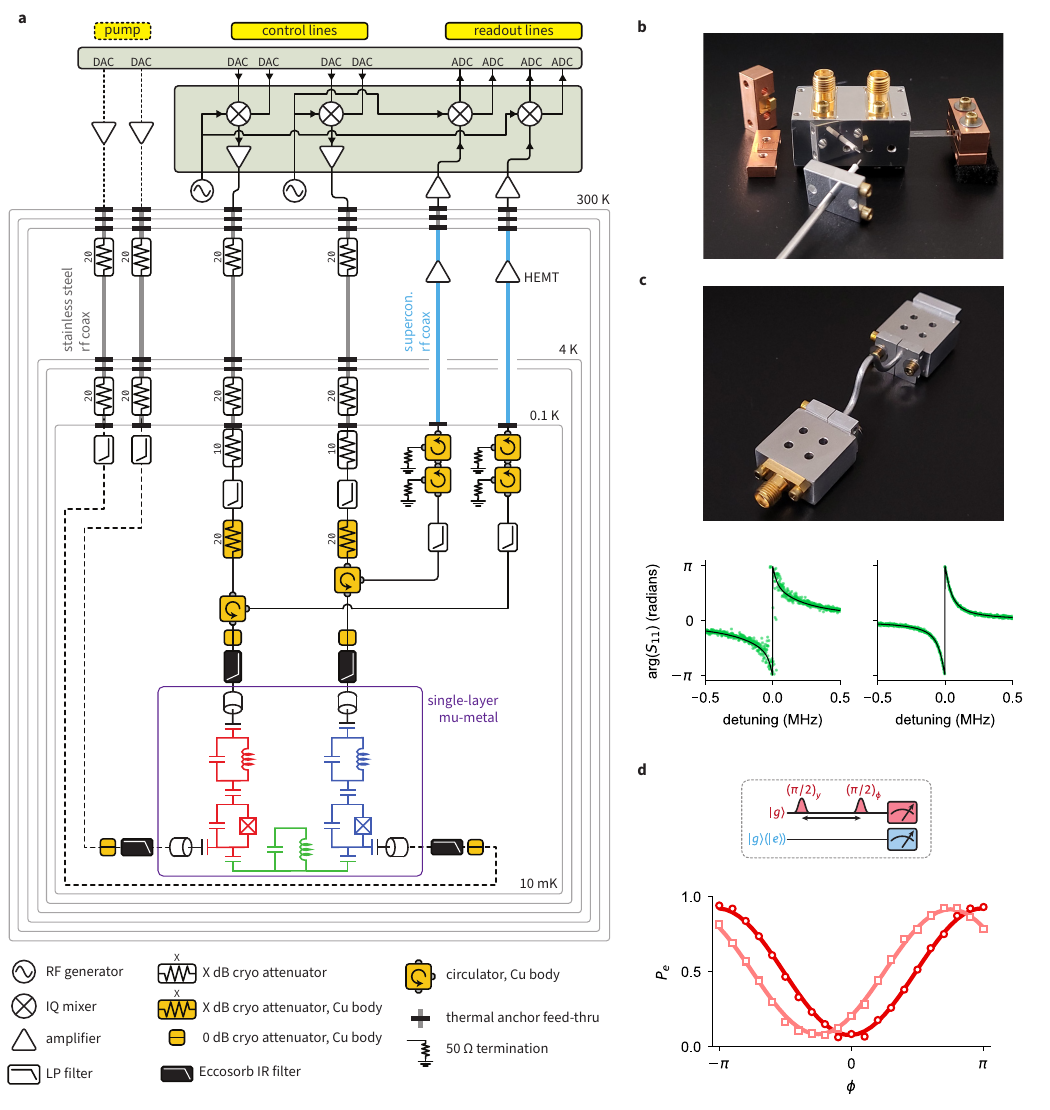}
	\caption{
    \textbf{Experimental setup.}
    \textbf{a,} Most of the wiring follows standard best practices \cite{krinnerEngineeringCryogenicSetups2019b}.
    Noteworthy components:
    \textbf{Low Pass (LP) filters:} probe lines: 12 GHz (\textit{K\&L 5L250-10200});
    pump lines: 1.8 GHz (\textit{Mini-circuits VLF-1800+}).
    \textbf{Output line configuration:} double-stage isolators (\textit{Low Noise Factory LNF-ISCIC4\_12A}), HEMT amplifier (\textit{Low Noise Factory LNF-LNC4\_8C});
    room temperature low-noise amplifier (\textit{Low Noise Factory LNF-LNR4\_14C}).
    \textbf{Room temperature electronics:} signal modulation and demodulation: \textit{Quantum Machines OPX+} and \textit{Quantum Machines Octave};
    external RF generators: \textit{SignalCore SC5511A};
    pump signal amplifier: \textit{Mini-circuits ZHL-1-2W-S+}.
    \textbf{b,} A single transmon module, based on \cite{axlineArchitectureIntegratingPlanar2016}.
    The chip is held from both sides of the tunnel by Cu clamps that are mounted to the package.
    Inside each clamp, a BeCu spring holds the chip in place with an estimated force of 250 grams.
    \textbf{c,} (Top) Setup used to measured the modes of the coaxial cable in reflection.
    The inner conductor of the cable sticks inside a tunnel in the square modules and couples to a pin connected to the SMA flange.
    (Bottom) Example reflection measurements of $3\lambda/2$ modes of two cables (left: NbTi cable; right: Al cable) obtained with a vector network analyzer; only phase response shown here.
    Fits (solid lines) to the data yield $Q_b = 3\times10^4$ (NbTi) and $5\times10^5$ (Al).
    \textbf{d,} Measurement of the direct dispersive interaction between the two qubits.
    We have measured Ramsey fringes on qubit~1, $\qty{4}{\micro\second}$ after qubit~2 was prepared in either $\ket{g}$ or $\ket{e}$.
    Solid lines: sinusoidal fits. 
    }
\label{fig:wiring diagram and device}
\end{figure*} 

\paragraph*{Cryogenic measurement setup}
Devices were cooled to 10\,mK in an \emph{Oxford Instruments Triton 500} dilution refrigerator. 
A schematic of setup and device wiring is shown in \Cref{fig:wiring diagram and device}a.

\paragraph*{Qubit devices} 
Qubit and readout resonator fabrication, as well as the basic package design follows \cite{axlineArchitectureIntegratingPlanar2016}.
An exploded view of a device is shown in \Cref{fig:wiring diagram and device}b.
Here, the body of the sub-cutoff waveguide enclosure is made of 6061 aluminum alloy.
Material choices and device fabrication limit qubit relaxation times to around $\qty{100}{\micro\second}$, obtained from simulations and control devices.

\paragraph*{Cable assembly and characterization}
Cables are clamped by Al brackets (see \cref{fig:fig1}b and \cref{fig:wiring diagram and device}b).
At the ends of the cable the outer conductor was removed with a razor blade to realize a capactive coupling between transmon pad and inner conductor \cite{burkhartErrorDetectedNetworking3D2020}.
To obtain good superconducting contact, brackets and device packages were mirror-polished on a granite block, submerged in soapy water, with up to 4000-grit sandpaper.
We characterized the assembly in a test setup that allows direct measurement of bus modes using a vector network analyzer (VNA), see \Cref{fig:wiring diagram and device}c.
Al cables used in the network were 2.2\,mm Al-PTFE-Al coaxial cable ordered from \textit{Nanjing HMC System Co}.
Initial tests on similar cables obtained from \textit{Microcoax} showed similar performance; we were not able to source these cables anymore after initial measurements.
In the network presented in the main text, the length of the cable is 73\,mm, with a free spectral range (FSR) of about 1.3\,GHz.
The cable was chosen to be short for economical reasons; at the initial project stage we faced difficulties sourcing suitable cables.
The only requirement for our experiment to work is that $\Omega$ is much smaller than the FSR.

\paragraph*{Network characterization}
System parameters for qubits, bus mode, and dispersive interactions are listed in \Cref{table: system parameters}.
The cross-Kerr between each qubit and the bus is found by measuring the qubit frequency before and after swapping an excitation into the bus \cite{burkhartErrorDetectedStateTransfer2021a}.
The cross-Kerr term $\chi_{a_1a_2}$ between the qubits shown in \Cref{table: system parameters} is found using Ramsey fringes (\cref{fig:wiring diagram and device}d).

\begin{table}[t]
\begin{center}
\begin{tabular}{ l c c c }
    \hline
    \hline
    Qubit $a_1$  & \ \ \ \ \ \ \ \ \ \ \ \ \ \ \ \ & & Value\\
    \hline
    Mode frequency & & $\omega_{a_1}/2\pi$ & $\qty{4675.5}{\mega\hertz}$ \\
    Relaxation time & & $T_1$ & $\qty{62 \pm 28}{\micro\second}$ \\
    Ramsey decay & & $T_{2\mathrm{R}}$ & $\qty{22 \pm 1}{\micro\second}$ \\
    Hahn echo decay & & $T_{2\mathrm{E}}$ & $\qty{34 \pm 2}{\micro\second}$ \\
    \hline
    Qubit $a_2$  &  &  & Value\\
    \hline
    Mode frequency & & $\omega_{a_2}/2\pi$ & $\qty{5510.7}{\mega\hertz}$ \\
    Relaxation time & & $T_1$ & $\qty{25 \pm 1}{\micro\second}$ \\
    Ramsey decay & & $T_{2\mathrm{R}}$ & $\qty{8 \pm 3}{\micro\second}$ \\
    Hahn echo decay & & $T_{2\mathrm{E}}$ & $\qty{20 \pm 1}{\micro\second}$ \\
    \hline
    Bus $b$ & & & Value\\
    \hline
    Mode frequency & & $\omega_b/2\pi$ & $\qty{5173.1}{\mega\hertz}$ \\
    Relaxation time & & $\tau_b$ & $\qty{6.2 \pm 0.2}{\micro\second}$ \\
    \hline
    Cross-Kerr & & & Value\\
    \hline
    Qubit $a_1$ and bus & & $\chi_{a_1b}/2\pi$ & $\qty{3.0}{\mega\hertz}$ \\
    Qubit $a_2$ and bus & & $\chi_{a_2b}/2\pi$ & $\qty{9.2}{\mega\hertz}$ \\
    Qubits $a_1$ and $a_2$ & & $\chi_{a_1a_2}/2\pi$ & $\qty{8.0}{\kilo\hertz}$ \\
    \hline
    \hline
\end{tabular}
\caption{
    \textbf{System parameters.}
    Mode frequencies,  lifetimes, and interaction strengths for the left qubit ($a_1$), right qubit ($a_2$), and bus mode ($b$).
    Uncertainties in the relaxation and coherence times reflect the fluctuations over the course of the experiment (about 3 months).
    Target $\chi$ values for qubit-cable coupling was $\qty{2}{\mega\hertz}$.
    Qubit coherence times were lower and fluctuate more strongly than in non-network packages.
    Typical $T_{1,2}$ times in qubits made in the same way but in packages without cable mount are reliably in the $\qtyrange{50}{100}{\micro\second}$ range, limited by fabrication and and sample holder material.
}
\label{table: system parameters}
\end{center}
\end{table}

\subsection{Data analysis}
\label{app: data analysis}

\paragraph*{SPAM errors}
The experiments reported were conducted without using a parametric amplifier, and thus relatively low state assignment fidelity ($P_{a_1}(g|g) = 98.5(5) \%$, $P_{a_1}(e|e) = 96.0(8) \%$; $P_{a_2}(g|g) = 99.3(3) \%$, $P_{a_2}(e|e) = 92.3(9) \%$).
For that reason, population data presented in the main text (with the exception of \cref{fig:fig3}b) were corrected for state preparation and measurement (SPAM) errors.
To perform the correction, each measurement is preceded by a calibration in which the qubits are measured after preparation in $\ket{g}$ and $\ket{e}$. 
We used a \emph{KMeans} clustering algorithm \cite{scikit-learn} to obtain misassignment probabilities, and then performed readout correction using that information \cite{gellerEfficientCorrectionMultiqubit2021, severinSuperconductingQubitReadout2023}.
To ensure good state preparation, all qubit gates were cheracterized regularly using interleaved randomized benchmarking (IRB). 
All single-qubit gates were tuned to achieve sub-percent errors; the errors obtained are consistent with being coherence-limited.
This IRB characterization ensures a small systematic error in population data.
We note, however, that the central figures of merit reported are obtained from measuring decay curves that are insensitive to SPAM.

\paragraph*{Uncertainties and error bars}
All population data shown are the mean readout signal after SPAM calibration.
Statistical uncertainty on the excited state population depends on the number of samples taken and was inferred by bootstrapping state assignment and readout correction.
For population data points without error bars, uncertainty is smaller than the marker size.
To place uncertainties on the figures or merit we have employed bootstrapping as well.
Raw data were randomly sub-sampled and analysis was performed for each subset.
Mean and uncertainty are obtained from the distribution of outcomes in the bootstrapped analysis.

\section{Sidebands}
\label{app: sidebands}

\begin{figure*}
\includegraphics{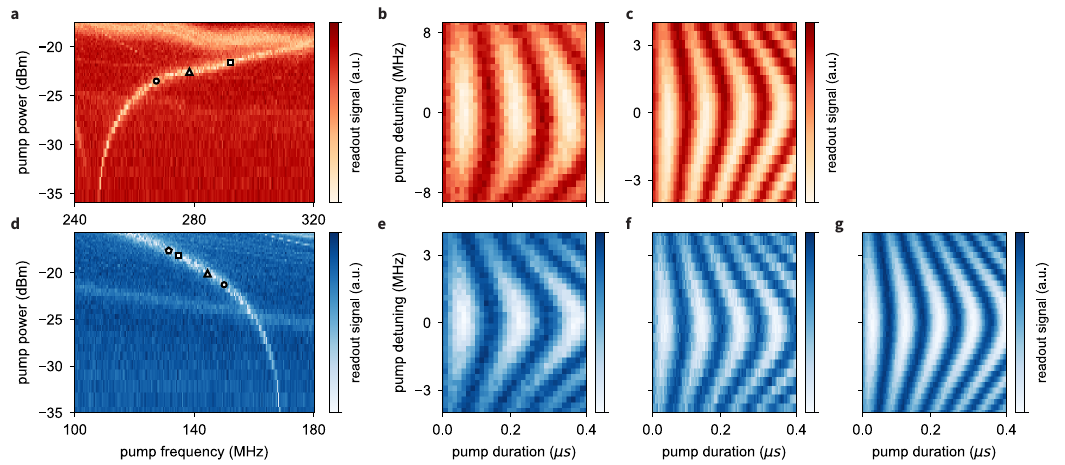}
	\caption{
    \textbf{Sideband spectroscopy and additional time-domain measurements.}
    \textbf{a,d,} Excited state population of qubits~1 (red) and 2 (blue) as a function of frequency and device input power of the microwave drives.
    Markers indicate where time-domain data was taken.
    Circles: $\Omega/2\pi \approx \qty{5}{\mega\hertz}$ (\cref{fig:fig2}c,d).
    Triangles: $\Omega/2\pi \approx \qty{7}{\mega\hertz}$ (\textbf{b,e}).
    Squares: $\Omega/2\pi \approx \qty{10}{\mega\hertz}$ (\textbf{c,f}).
    Pentagon: $\Omega/2\pi \approx \qty{11}{\mega\hertz}$ (\textbf{g}).
    Qubit~1 displays additional transition(s) crossing the sideband resonance near the triangle marker, resulting in a broadened and asymmetric oscillation pattern (note the different scale on the detuning axis in \textbf{b}).
    }
\label{fig:de-excitation plots}
\end{figure*}

\subsection{Theoretical model}
\label{app: sideband model}

The dispersive Hamiltonian for a transmon ($\hat{a}$) coupled to a microwave resonator ($\hat{b}$) is given by (setting $\hbar \equiv 1$)
\begin{equation}
H = \omega_a \hat{a}^\dag \hat{a} + \omega_b \hat{b}^\dag \hat{b} - E_{\mathrm{J}}\left(\mathrm{cos}(\varphi) + \frac{\varphi^2}{2}\right) 
\label{eq: transmon and resonator}
\end{equation}
where $\omega_a$ and $\omega_b$ are the transmon and resonator frequencies, respectively.
The phase across the Josephson junction $\varphi = \varphi_a (\hat{a}^\dag + \hat{a}) + \varphi_b (\hat{b}^\dag + \hat{b})$ contains contributions from both the transmon and resonator modes, allowing for wave mixing. 
We then apply two microwave pump tones to the transmon, described by the additional Hamiltonian term:
\begin{equation}
H_{\mathrm{pump}} = \sum_{i=1}^{2} \epsilon_i (\hat{a}^\dag e^{-i \omega_i t} + \hat{a} e^{i \omega_i t})
\label{eq: pump hamiltonian}
\end{equation} 
where $\epsilon_i$ is the amplitude of the pump tone and $\omega_i$ is its frequency.
Following the standard procedure (see, e.g., \cite{campagne-ibarcqDeterministicRemoteEntanglement2018b}), we move to a displaced frame for each pump tone, with displacements $\xi_i  = \left(\epsilon_i e^{-i \omega_i t}\right)/\left(\omega_a - \omega_i\right)$, and to a rotating frame for both modes $\hat{a}$ and $\hat{b}$.
After expanding the transformed Hamiltonian up to fourth order we can isolate the swap term $\propto \hat{a}^\dag \hat{b}$ with coefficient
\begin{multline}
    \Omega/2 = E_{\mathrm{J}} \varphi_a^2 \varphi_b^2 e^{i(\omega_a - \omega_b)t} \\
    \times \left((\xi_1^2 + \xi_1^{*2})/2 + (\xi_2^2 + \xi_2^{*2})/2 + 
    \xi_1^* \xi_2 + \xi_2^* \xi_1\right).
    \label{eq: swap coeff}
\end{multline}
This term averages out in the rotating wave approximation (RWA), unless the frequency matching condition $\omega_2 \pm \omega_1 = |\omega_a - \omega_b|$ is satisfied. 
Having $\omega_2 - \omega_1 = |\omega_a - \omega_b|$ results in the $\xi_1^* \xi_2 + \xi_2^* \xi_1$ terms surviving the RWA. 
However, having $\omega_2 + \omega_1 = |\omega_a - \omega_b|$ such that $\omega_1 = \omega_2 = (\omega_a - \omega_b)/2$ results in all terms in \Cref{eq: swap coeff} surviving the RWA, increasing the coefficient by a factor of two. 
In other words, for the degenerate case we expect twice the sideband rate for a fixed number of pump photons compared to the nondegenerate case. 
Given that increasing pump power results in detrimental effects \cite{shillitoDynamicsTransmonIonization2022}, achieving the desired dynamics with the lowest amount of power is highly desireable.
This is a primary motivation for using the low frequency parametric pumps employed in the experiment.

\subsection{Tuneup and limitations}
\label{app: sideband tuneup}

Sideband oscillations for different pump powers are shown in \Cref{fig:de-excitation plots}.
The resonances shift as a function of applied power due to Stark shifts on qubit and bus, $-\delta_a \hat{a}^\dagger \hat{a}$ and $-\delta_b \hat{b}^\dagger \hat{b}$ respectively, where $\delta_i$ is proportional to the pump power applied.
While for qubit~2 `clean' oscillations above 10\,MHz can be induced, we observe distorted oscillations above 5\,MHz for qubit~1.
At this moment we have no explanation for the features besides the sideband resonance that appear to cause the qubit state to change.
It is an outstanding question whether additional degrees of freedom (such as coupling to uncontrolled TLS) are needed to explain these features, or whether they are a result from strong-driving behavior of the transmon itself \cite{shillitoDynamicsTransmonIonization2022}.

\section{Interchip gates}
\label{app: interchip gates}

\begin{figure}
\includegraphics{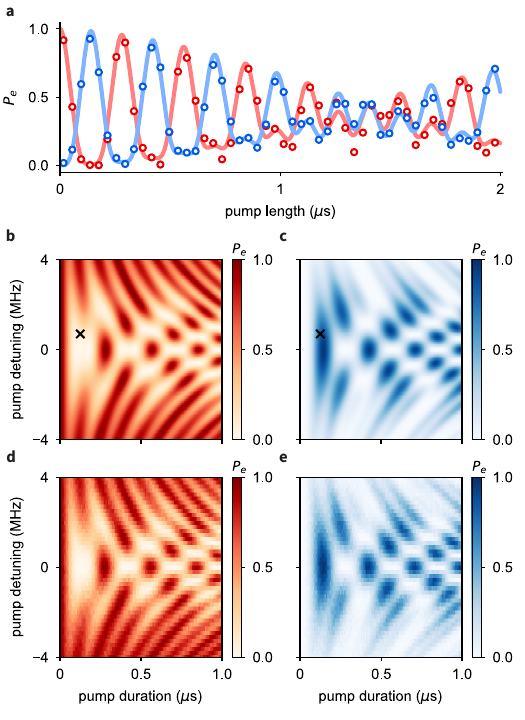}
	\caption{
    \textbf{Time-dynamics of resonant Raman transitions.}
    \textbf{a,} Raman transitions between the qubits~$a_1$ (red) and $a_2$ (blue), immediately after tuning both sidebands individually to 5\,MHz.  
    The simulation (solid lines) matches the data (circles) with a $\qty{0.36}{\mega\hertz}$ detuning term on the bus (\cref{eq: detuned swap hamiltonian}).
    \textbf{b,c,} Simulation of the excited state population for qubits~1 (red) and 2 (blue) as a function of the drive detuning and time, assuming tuned up sidebands and no bus detuning.
    \textbf{d,e,} Data for the detuned swaps for the left and right qubit.
    In the simulation plots, `X' marks a point of interest: 
    in ref.~\cite{burkhartErrorDetectedStateTransfer2021a} this point corresponds to an entangling gate.
    Here, we only predict a concurrence of 0.91 at this point, lower than what can be achieved using the stroboscopic approach shown in \cref{fig:fig3}e of the main text.
    }
    
\label{fig:qubit-qubit dynamics}
\end{figure}

\subsection{Prediction of SWAP efficiency}
\label{app: swapp efficiency}

Here we describe how the calculations resulting in \Cref{fig:fig1}d were performed.
We consider the general case with the Raman drives being detuned by $\Delta$,
\begin{equation}
    H = \dfrac{\Omega}{2} \Bigl(\hat{a}_1^\dagger \hat{b} + \hat{a}_1\hat{b}^\dagger \Bigr) + \dfrac{\Omega}{2} \Bigl(\hat{a}_2^\dagger \hat{b} + \hat{a}_2\hat{b}^\dagger \Bigr) + \Delta \hat{b}^\dagger \hat{b}.
    \label{eq: detuned swap hamiltonian}
\end{equation}
Increasing the detuning increases the time it takes to fully swap an excitation between the qubits.
This swap time can be found by considering the beam splitter angle, which measures how much of an excitation is transferred between qubits while leaving the bus unpopulated after the beam splitter \cite{burkhartErrorDetectedNetworking3D2020}.
The beam splitter angle is
\begin{equation}
    \theta = \dfrac{\pi}{2} \biggl(1 - \dfrac{\Delta}{\sqrt{8(\Omega/2)^2 + \Delta^2}} \biggr),
    \label{eq: bs angle}
\end{equation}
where $\theta = \pi/2$ corresponds to a full excitation swap.
$\Omega$ is the usual sideband Rabi rate between each qubit and the bus.
Next, the beam splitter time ($\tau_{\mathrm{BS}}$), the time between instances when there is no photon population in the bus, can be written as
\begin{equation}
    \tau_{\mathrm{BS}} = \dfrac{2\pi}{\sqrt{8(\Omega/2)^2 + \Delta^2}}.
    \label{eq: bs time}
\end{equation}
To perform a full excitation swap, one can execute $N$ beam splitter operations with angle $\theta = \pi/2N$, where $N$ is a positive integer $\geq 1$.
Note that it follows that for a given $\Omega$, only certain $\Delta$ can be used to execute a full swap.
The swap time can be determined as $\tau_{\mathrm{SWAP}} = N\tau_{\mathrm{BS}}$.
To obtain valid swap times in \Cref{fig:fig1}d we have fixed $\Omega$, and then for each $N$ found (the unique) $\Delta$ and $\tau_{\mathrm{SWAP}}$ for a single excitation swap.
From there, the swap efficiency is determined by numerically evaluating the Lindblad master equation using the Hamiltonian in \cref{eq: detuned swap hamiltonian} for a given swap time, bus quality factor $Q_b$, and sideband Rabi rate $\Omega$.

\subsection{Resonant Raman transitions}
\label{app: resonant Raman}

\paragraph*{Tuneup of continuous evolution}
Here we describe how the system was tuned to obtain the data in \Cref{fig:fig3}a.
Each sideband is first tuned separately to a specified value of $\Omega$ (here, 5\,MHz).
When both pumps are on, the total Stark shift is different compared to individual pumps, creating a mismatch in Rabi rates.    
\Cref{fig:qubit-qubit dynamics}a shows continuous evolution after each sideband was only tuned individually.
To restore matching we perform an iterative procedure to match the evolution to the predicted one.
We evaluate the RMS error of mismatch between data and expected ground/excited state populations for both qubits at multiples of the expected swap time, $\tau_\mathrm{SWAP} = \sqrt{2}\pi/\Omega$.
This error is minimized by iteratively tuning both pump frequencies and amplitudes.
Tuneup is finished when the error converges near a minimum error or if the error is smaller than a preset threshold.
Successful tuneup can be confirmed by measuring the full time evolution of both qubit state populations as a function of the drive detuning and comparing to simulation (\cref{fig:qubit-qubit dynamics}b-e).

\paragraph*{Analysis of continuous Raman transitions}
To determine the loss per swap when using continuous drives, we find the excited state probability $P_e$ by fitting to the following equation:

\begin{equation}
    P_e = P_e(0) \Bigl[ e^{-t/\tau} \cos \Bigl(\dfrac{\tilde{\Omega}t}{2\sqrt{2}} + \phi \Bigr) + \dfrac{1}{2}\Bigl(1 - e^{-t/\tau}\Bigr) \Bigr]^2,
    \label{eq: qubit swap population}
\end{equation}
where
\begin{equation}
    \tilde{\Omega} = \Omega \sqrt{1 - \Bigl(\dfrac{2}{\sqrt{2}\Omega \tau} \Bigr)^2}.
    \label{eq: effective omega}
\end{equation}
$\Omega$ is the sideband Rabi rate, $\phi$ a phase offset, and $\tau$ is the decay time of the envelope for the swaps \cite{burkhartErrorDetectedNetworking3D2020}.
In an ideal system with matched Rabi rates and loss only coming from the bus, $\tau$ should be twice the bus lifetime.
The swap time is then divided by the decay time $\tau$ to give us the percent loss per swap.
From fitting the data to \cref{eq: qubit swap population}, for qubit $a_1$, $\Omega/2\pi = \qty{5.040 \pm 0.001}{\mega\hertz}$ and $\tau = \qty{14.2 \pm 0.7}{\micro\second}$.
For qubit $a_2$, $\Omega/2\pi = \qty{5.041 \pm 0.002}{\mega\hertz}$, and $\tau = \qty{13.3 \pm 0.4}{\micro\second}$.
For a Rabi rate of 5.04\,MHz the swap time is 140.3\,ns.
To verify these results and to obtain a best estimate for our loss per swap, we have simulated the time dynamics, factoring in the measured coherence times.
From simulation we obtain a decay time of $\qty{12.4 \pm 0.4}{\micro\second}$, twice the average bus lifetime recorded in \Cref{table: system parameters}.
This is the solid curve shown in \Cref{fig:fig3}a, in very good agreement with experimental data.
From this simulation, the loss per swap is 0.97(2)\%.

\begin{figure}
\includegraphics{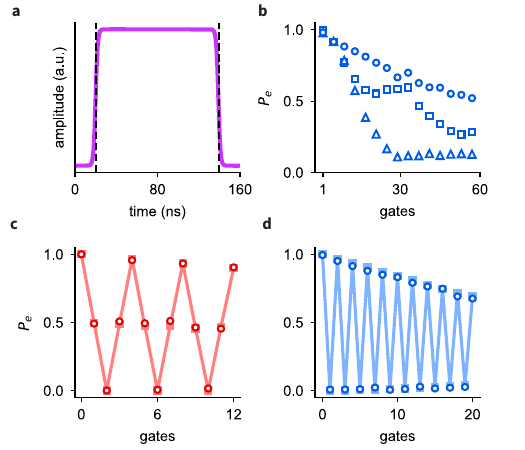}
	\caption{
    \textbf{Tuning up gates between qubits.}
    \textbf{a,} Example of a smooth pulse used to generate the swap gate.  
    For this pulse shown, $T = \qty{160}{\nano\second}$, $t_{\mathrm{eff}} = \qty{120}{\nano\second}$, and $\sigma = \qty{4}{\nano\second}$.
    Blacked dashed lines represent the effective gate time length.
    \textbf{b,} Excited state population for the right qubit as a function of consecutive gates for different effective gate times.
    Triangles, squares, and circles correspond to effective gate times of 139, 139.5, and 141.5\,ns, respectively. 
    Here, the 141.5\,ns gate is best, as the exponential decay is indicative of least residual coherent error.
    \textbf{c,d,} excited state population for qubit $a_1$ (red) and $a_2$ (blue) as a function of the sideband gates applied where $\Omega/2\pi \approx \qty{5}{\mega\hertz}$.
    Circles are data, squares are from simulation.
    Half-swap pulses are tuned between qubit~$a_1$ and bus, full-swap pulses between bus and qubit~$a_2$.
    The optimal effective sideband gate times for the left and right qubits are 52.6\,ns and 101.8\,ns, respectively.
    }
\label{fig:qubit-qubit gate tune up}
\end{figure}

\paragraph*{Tuneup of pulsed swaps}
As described above, swap time is inferred from analysis of the continous evolution.
The simplest way to apply a SWAP gate is to play a rectangular pulse for this time.
However, smoothed pulse edges are required to constrain the pulse bandwidth.
We have shaped gate pulses as (\cref{fig:qubit-qubit gate tune up}a)
\begin{equation}
    \Bigl[1 + \tanh \Bigl(\dfrac{2(t-\tilde{t})}{\sigma} \Bigr) \tanh \Bigl(\dfrac{2(-t + (T - \tilde{t}))}{\sigma} \Bigr) \Bigr]/2 
    \label{eq: shaped pulse}
\end{equation}
where $\tilde{t} = (T - t_{\mathrm{eff}})/2$.
$T$ is the total length given for the pulse, and $\sigma$ is half the total time of the smooth edges on either side of the flattop pulse.
$t_{\mathrm{eff}}$ is the effective length of the pulse such that it has the same area under it as a rectangular pulse with the same length and amplitude.
Optimization of the gate is achieved by varying $t_{\mathrm{eff}}$.

\paragraph*{Stroboscopic entanglement tune up} 
To tune up the sideband gate pulses used for entanglement, first each sideband is set to a desired Rabi rate $\Omega$, and then the swap time into the bus is found simply by $2\pi/\Omega$.
From there, we shape the sideband gates as smooth, flattop pulses (\cref{fig:qubit-qubit gate tune up}a) and begin tuning over the effective sideband gate time to find the highest fidelity swap into the bus.
This can be done by looking at the excited state population after a train of sideband pulses (\cref{fig:qubit-qubit gate tune up}c,d).
The error per swap is minimized through the effective gate time.
To quantify the error we take the RMS error of deviation from the ideal population (ground or excited state, alternatingly) and fit to an exponential decay.

\subsection{Detuned Raman transitions}
\label{app: detuned Raman}

\paragraph*{Large detuning model}
Treating both qubits in the frame rotating at the frequency of their corresponding pump tones, the Hamiltonian is
\begin{equation}
    H_{\text{sys}} = \Delta \hat{a}^{\dagger}_1 \hat{a}_1 + (\Delta - \delta) \hat{a}^{\dagger}_2 \hat{a}_2 + \frac{\Omega}{2}( \hat{a}^{\dagger}_1 \hat{b} + \hat{a}^{\dagger}_2 \hat{b} + \text{h.c.}).
\end{equation}
To simplify this Hamiltonian, we apply a Schrieffer-Wolff transformation \cite{blaisCircuitQuantumElectrodynamics2021d}.
Since $\Omega \ll \Delta$, we treat the mode couplings as a perturbation, and write the Hamiltonian as:
\begin{equation}
    H_{\text{sys}} = H^0_{\text{sys}} + V,
\end{equation}
where $H_{\text{sys}}^0 = \Delta \hat{a}^{\dagger}_1 \hat{a}_1 + (\Delta - \delta) \hat{a}^{\dagger}_2 \hat{a}_2 $ and $V =  \frac{\Omega}{2}(\hat{a}^{\dagger}_1 \hat{b} + \hat{a}^{\dagger}_2 \hat{b} + \text{h.c.})$. 
Up to the first order in $V$, the transformed Hamiltonian in the dressed basis is now:
\begin{align}
    \Bar{H}_{\text{sys}} &= \Delta(1 + \frac{\Omega^2}{4 \Delta^2}) \hat{a}^{\dagger}_1 \hat{a}_1 + (\Delta - \delta) (1 + \frac{\Omega^2}{4 (\Delta - \delta)^2})\hat{a}^{\dagger}_2 \hat{a}_2 \nonumber \\ 
    &+ \frac{\Omega^2}{2} \left( \frac{1}{2 \Delta} + \frac{1}{2 (\Delta - \delta)} \right) ( - \hat{b}^{\dagger} \hat{b} + \hat{a}^{\dagger}_1 \hat{a}_2 + \hat{a}_1 \hat{a}^{\dagger}_2).
\end{align}
Given that $\Delta~\text{and}~(\Delta - \delta) \gg \Omega$, we can omit the linear terms of the system.

\paragraph*{Dual-rail qubit tune up}

We implement the detuned Raman process near $5$~MHz qubit-swap rate.
We detune the pump on qubit $a_1$ by 20\,MHz ($\Delta/2\pi = \qty{20}{\mega\hertz}$), and perform two-tone spectroscopy on qubit $a_1$ while sweeping the pump frequency of qubit $a_2$ until an avoided crossing pattern in found.
The initialization of the logical qubit ($|gg\rangle \rightarrow |0\rangle{_\text{L}}~\text{or}~|1\rangle_\text{L}$) is realized by local single-qubit gates, while its manipulation is implemented via detuned Raman drives.
The dual-rail qubit $\pi$ pulse is tuned up by finding the pump duration that fully swaps the excitation from one physical qubit to another. 
To do so, we first initialize the dual-rail qubit in the $|0\rangle_\text{L}$ state, then turn on both of the pumps at the frequency matched point (i.e., $\delta = 0$) while sweeping the pump duration. 
We fit the error (residual $|0\rangle_\text{L}$ population) near the full swap point to a quadratic model and identify the optimal duration.
The $\pi/2$ pulse is implemented by applying pumps with half the length of the $\pi$ pulse.

\paragraph*{Frame evolution}
The dual-rail qubit will leave the rotating frame when no pulses are applied, and will accumulate a relative phase between $|0\rangle{_\text{L}}$ and $|1\rangle{_\text{L}}$ at the rate of the pump frequency detuning $\Delta$.
The Echo measurement allows for extracting this detuning quantitatively, and specifically in our case, $\Delta/2\pi = \qty{31.43}{\mega\hertz}$.
This deviates from the initial setting of 20\,MHz, which we attribute to the pump induced Stark shift.
We can estimate the qubit-bus coupling strength by applying this value to the model that fits the time evolution data (\cref{fig:fig4}c); we get $\Omega / 2\pi = \sqrt{\Omega_\text{R} \Delta} / 2\pi \simeq \qty{6.6}{\mega\hertz}$, which is close to our starting point, 5\,MHz.
We attribute this difference to the variations in the effective parametric pumping strength at different pump frequencies.

\section{Repeatability}
\label{app: repeatability}

\begin{figure}
\includegraphics{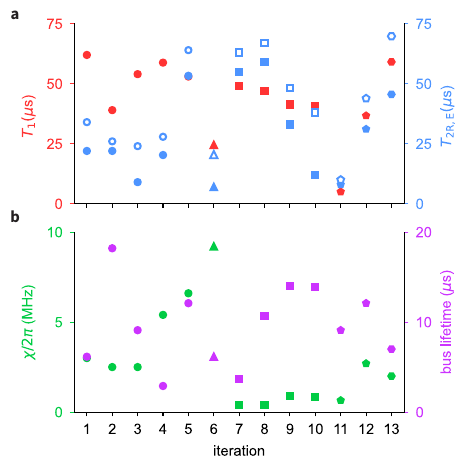}
	\caption{
    \textbf{Repeatability measurements.}
    Each iteration is a separate cycle of a bus re-mounted to a qubit package, cooled down and measured. 
    Exception is iteration 8, where the device was not disassembled and reassembled from iteration 7, but only thermally cycled.
    Different markers represent different qubit samples and packages.
    Together, iterations 1 and 6 are the ones during which the main experiment data was collected.
    We have obtained `clean' Sideband oscillations with $\Omega/2\pi \geq \qty{5}{\mega\hertz}$ in all iterations but iterations 7 and 8; there, we achieved only $\Omega/2\pi \approx \qty{4.5}{\mega\hertz}$ before observing high-power effects.
    \textbf{a,} Relaxation times (solid red), Ramsey (solid blue), and echo decay times (open blue).
    \textbf{b,} Qubit-bus dispersive shift $\chi$ (green) and bus lifetime (purple) across each iteration.
    }
\label{fig:repeatability test}
\end{figure}

\Cref{fig:repeatability test} shows qubit coherence times, dispersive shifts between the qubit and bus, and the bus lifetimes for a total of 13 different mounting cycles across five different samples.
Cycles consisted of mounting the bus to a package containing a qubit, cooling to base temperature and performing measurements, warming up, removing and re-attaching the bus.
Importantly, we were able to reach $\Omega/2\pi \geq \qty{5}{\mega\hertz}$ sidebands in 11 out of these 13 iterations. 
This indicates that while our mounting system is simple, it already allows for a reliable interconnect.
The wide spread of system parameters across these cycles is, however, noteworthy and points toward the need for improved mechanical designs.
In particular, coherence times are shorter and vary more than typical devices without the mounted cable in our lab; typical coherence times in similar packages but without cable mounts reach routinely \qtyrange{50}{100}{\micro\second}, limited by the single-step fabrication process used here and by the material of the qubit packages.
We believe that the mechanics of the mount need to be improved before state-of-the-art coherence in these types of packages \cite{ganjamSurpassingMillisecondCoherence2024} can be reached in combination with cable mounts.
As for the dispersive interaction between the transmon and cable bus mode, the design value is $\chi/2\pi = \qty{2}{\mega\hertz}$, but the measured spread is 0.4~MHz to 9.2~MHz.
From finite-element simulations we have found that this can be explained by a variation of inner-conductor to chip distance up $\pm$ 0.5\,mm from the target position.
This variation is, given the softness of Al cable, a realistic run-to-run spread.

\bibliography{references}
\end{document}